\documentstyle[tighten,preprint,aps]{revtex}

\begin{document}
\title{\boldmath Consistency of parity-violating pion-nucleon
  couplings extracted from measurements in $^{18}$F and $^{133}$Cs}
\author{W.S.~Wilburn and J.D.~Bowman} \address{Los Alamos National
  Laboratory, Los Alamos, NM 87545} \date{\today} \maketitle

\begin{abstract}
  The recent measurement of the nuclear anapole moment of $^{133}$Cs
  has been interpreted to yield a value of the weak pion-nucleon
  coupling $H_{\pi}^{1}$ which contradicts the upper limit from the
  $^{18}$F experiments. We argue that because of the sensitivity of
  the anapole moment to $H_{\rho}^{0}$ in the odd proton nucleus
  $^{133}$Cs, there is a combination of weak meson-nucleon couplings
  which satisfies both experiments and which is (barely) in agreement
  with theory. In addition, the anapole moment measurement in
  $^{205}$Tl gives a constraint which is inconsistent with the value
  from $^{133}$Cs, calling into question the theory of nuclear anapole
  moments. We argue that measurements of directional asymmetry in
  $\vec{n}+p\rightarrow d+\gamma$ and in the photo-disintegration of
  the deuteron by circularly polarized photons, combined with results
  from \emph{pp} scattering, would determine $H_{\pi}^{1}$ and several
  other weak meson-nucleon couplings in a model-independent way.
\end{abstract}

\section{Introduction}
  \label{sec:int}
  
  The nucleon-nucleon (\emph{NN}) weak interaction can be described as
  arising from meson exchange involving a parity-violating weak vertex
  and a parity-conserving strong vertex. This interaction leads to a
  meson-exchange potential involving seven weak meson-nucleon coupling
  constants\footnote{We use the notation of Adelberger and Haxton,
    where, for example $H_{\pi}^{1}=F_{\pi}=g_{\pi}f_{\pi}/\sqrt{32}$
    and $H_{\rho}^{0}=F_{0}=-g_{\rho}h_{\rho}^{0}/2$.}, of which two,
  the $\Delta I=1$ $\pi$\emph{N} coupling $H_{\pi}^{1}$ and the
  $\Delta I=0$ $\rho$\emph{N} coupling $H_{\rho}^{0}$ contribute the
  most to low-energy nuclear observables. The coupling $H_{\pi}^{1}$
  is of interest because it is expected to be particularly sensitive
  to weak neutral currents. In the absence of neutral currents,
  $H_{\pi}^{1}$ is small. Some theories that include neutral currents
  predict large values of $H_{\pi}^{1}$. Despite considerable
  experimental and theoretical effort, the size of the weak
  pion-nucleon coupling constant $H_{\pi}^{1}$ remains uncertain.
  Until recently, the conflict has centered around a discrepancy
  between theory, which predicts a relatively large value
  $H_{\pi}^{1}=1.1\times10^{-6}$ \cite{Des80}, and experiments in
  $^{18}$F, which set an upper limit of
  $H_{\pi}^{1}\leq0.28\times10^{-6}$ \cite{Ade85}, a disagreement of
  more than $3\sigma$. In addition, Adelberger and Haxton \cite{Ade85}
  fit the available data to obtain $H_{\pi}^{1}=0.5\times10^{-6}$, a
  value somewhat larger than the $^{18}$F upper limit. The
  interpretation of the $^{18}$F experiments is thought to be more
  reliable than for other nuclei \cite{Hax81}.  Although experiments
  in other nuclei have been used to determine the weak meson-nucleon
  couplings (see \cite{Hae95,Ade85} for reviews), their interpretation
  is considerably more model-dependent than $^{18}$F since their
  interpretation involves \emph{ab initio} calculations of transition
  matrix elements of the meson-exchange potential. For this reason,
  they will not be considered here.
  
  Recent progress has deepened the controversy. First we consider
  progress in the theory.  Kaplan and Savage \cite{Kap93} find that
  strangeness-changing currents, which were previously neglected,
  contribute substantially to $H_{\pi}^{1}$. Their result,
  $H_{\pi}^{1}=1.2\times10^{-6}$, is in agreement with the
  Desplanques, Donoghue, and Holstein (DDH) ``best'' value prediction.
  In contrast, Henley, Hwang, and Kisslinger \cite{Hen96}, using QCD
  sum rule techniques, find an almost complete cancellation between
  perturbative and non-perturbative contributions from weak neutral
  currents. This cancellation results in an extremely small
  $H_{\pi}^{1}=0.05\times10^{-6}$, a factor of 24 smaller than Kaplan
  and Savage and DDH\@.
  
  The interpretation of new data on the nuclear anapole moment of
  $^{133}$Cs has added a new dimension to the controversy.  The
  non-zero measurement of the anapole moment of $^{133}$Cs
  \cite{Woo97} has been analyzed by Flambaum and Murray \cite{Fla97}
  to extract a value for $H_{\pi}^{1}$. Their result,
  $H_{\pi}^{1}=2.26\pm0.50$(expt)$\pm0.83$(theor)$\times10^{-6}$ is a
  factor of two larger than the DDH value and a factor of \emph{seven}
  larger than the upper limit set by the $^{18}$F experiments.
  
  We argue that the because of the sensitivity of the anapole moment
  to $H_{\rho}^{0}$, the two experimental results are not
  incompatible.  However, the constraint on the weak couplings
  extracted from the $^{133}$Cs result is inconsistent with that
  extracted from an earlier null measurement of the anapole moment of
  $^{205}$Tl \cite{Vet95}, calling into question the reliability of
  the theoretical interpretation of the anapole moment. The issues
  raised by the apparent inconsistencies between different
  determinations of $H_{\pi}^{1}$ point out the need for a
  model-independent determination of the weak meson-nucleon couplings.

\section{\boldmath Extraction of Weak Couplings from $^{18}$F and
  Anapole Moment Experiments}
  \label{sec:18f}
  
  The determination of the weak-meson-nucleon couplings from
  experimental measurements in nuclei are discussed in the review by
  Adelberger and Haxton \cite{Ade85}. There are substantial
  uncertainties in interpreting most experiments in nuclei because one
  cannot make reliable \emph{ab initio} calculations of amplitudes of
  the weak-meson-exchange potential operators. The experiment to
  measure the circular polarization $P_{\gamma}$ of the 1081~keV
  transition in $^{18}$F is an exception to this unfortunate situation
  because the matrix elements needed to extract a value for
  $H_{\pi}^{1}$ from experiments can be measured. The circular
  polarization of a $\Delta I=1$ parity-forbidden gamma transition in
  $^{18}$F has been measured in five different and internally
  consistent experiments (references given in \cite{Ade85}). To a good
  approximation the circular polarization is due to the
  parity-violating mixing between the $J=0$, $I=0$ parity-odd level
  ($|-\rangle$) in $^{18}$F and the nearly degenerate $J=0$,
  even-parity, $I=1$ level ($|+\rangle$).  The circular polarization
  is given by:
  \begin{equation}
    P_{\gamma}=\frac{2}{\Delta E}\frac{\langle+|V_{pnc}|-\rangle
      \langle gs|M1|+\rangle}{\langle gs|E1|-\rangle}.
  \end{equation}
  The magnitudes of the $M1$ and $E1$ transition amplitudes and the
  energy splitting $\Delta E=39$~keV between the levels are known
  experimentally.  Bennet, Lowry, and Krien \cite{Ben81} and Haxton
  \cite{Hax81} point out that the unknown amplitude,
  $\langle+|V_{pnc}|-\rangle$, could be related to the lifetime of the
  first forbidden beta decay between isobaric analog of $|+\rangle$ in
  $^{18}$Ne and $|-\rangle$.  For beta transitions between opposite
  parity $J=0$ levels, spin and parity selection rules exclude all but
  two of the six possible transition amplitudes. One of these vanishes
  in the long wavelength limit leaving only the $M_{00}^{5}$
  amplitude.  Haxton use PCAC, current algebra, and the approximation
  that the neutron and proton densities are the same for the mass 18
  system to argue that the two-body part of $M_{00}^{5}$ renormalizes
  the one body part and that the operator for the weak nucleon-nucleon
  potential due to pion exchange is to within a known constant an
  isospin rotation of the operator for $M_{00}^{5}$. Haxton evaluates
  the contributions of $\rho$ and $\omega$ to be a 5\% correction to
  the pion term. Since the experimental value,
  $P_{\gamma}=8\pm39\times10^{-5}$, of the asymmetry is consistent
  with zero, an upper limit for $H_{\pi}^{1}$ results. Adelberger and
  Haxton find $H_{\pi}^{1}\leq0.28\times10^{-6}$.  This value is a
  fraction of the DDH ``best'' value and is an order of magnitude
  smaller than the ``reasonable range''. It should be noted, however,
  that Kaplan and Savage have suggested that including two-pion
  contributions may reduce the sensitivity of $P_{\gamma}$ to
  $H_{\pi}^{1}$ through interference with the one-pion term
  \cite{Kap93}.
  
  The anapole moment operator \cite{Zel58} is a parity-odd rank one
  tensor and is given by
  \begin{equation}
    \vec{a}=-\pi\int r^{2}\vec{\jmath}(r)\,d^{3}r,
  \end{equation}
  where $\vec{\jmath}(r)$ is the electromagnetic current density
  operator. The anapole moment has an expectation value of zero for
  states of definite parity. The size of the anapole moment is given
  by the dimensionless constant $\kappa_{a}$, whose value is thought
  to be theoretically stable for three reasons:
  \begin{enumerate}
  \item The anapole moment of a nuclear state is a diagonal matrix
    element, in contrast to a transition matrix element.
  \item The nuclear wave function is constrained by the measured
    magnetic moment of the state. The single-particle estimate is
    $\mu_{sp}=1.72$ for $^{133}$Cs and $\mu_{sp}=2.79$ for $^{205}$Tl
    (see, for example, \cite{Boh69}) in relatively good agreement with
    the experimental values $\mu_{exp}=2.58$ and $\mu_{exp}=1.64$,
    respectively.
  \item The contribution to anapole moment of the spin of the odd
    proton can be estimated analytically and gives results close to
    those from full many-body calculations\footnote{Dmitriev and
      Telitsin find that a naive harmonic oscillator model gives
      results which agree with their full calculation to within 10\%.}
    \cite{Dmi97}.
  \end{enumerate}
  The estimate of the one-body part of the anapole moment is similar
  to the calculation of the contribution of the spin of the unpaired
  nucleon to the magnetic moment of a nucleus.  The analytical
  estimate \cite{Fla80} gives a value in rough agreement with detailed
  nuclear structure calculations by Haxton \cite{Hax97}, Haxton,
  Henley, and Musolf \cite{Hax89}, Flambaum and Murray \cite{Fla97},
  and Dmitriev and Telitsin \cite{Dmi97}. The latter is the most
  recent calculation and includes the effects of spin, spin-orbit,
  convection, and contact currents, many-body corrections, and RPA
  re-normalization of the weak interaction addition to the
  single-particle weak interaction. An additional attractive feature
  of anapole moment measurements is that they can in principle be made
  for many nuclei, thus providing an opportunity to test the
  consistency of the theory.

\section{\boldmath Interpretation of the $^{133}$Cs Anapole Moment Measurement}
  \label{sec:csa}
  
  Flambaum and Murray have argued that the value of $H_{\pi}^{1}$
  determined from the $^{18}$F measurement and the measured value of
  the $^{133}$Cs anapole moment are inconsistent. As discussed in the
  introduction, they argue that the $^{133}$Cs result requires a much
  larger value of $H_{\pi}^{1}$ than the $^{18}$F result. The
  $^{133}$Cs anapole moment is, however, almost as sensitive to
  $H_{\rho}^{1}$ as to $H_{\pi}^{1}$, as we demonstrate below. Because
  of this sensitivity, and because of the lack of model-independent
  constraints on $H_{\rho}^{0}$, it possible to extract values of the
  two couplings which agree with both experiments.
  
  We follow the method of Flambaum and Murray in relating the
  $^{133}$Cs anapole moment to the weak couplings. Dmitriev and
  Telitsin \cite{Dmi97} calculate for the case of $^{133}$Cs
  \begin{equation}
    \kappa_{a}=0.041g_{p}+0.008g_{n}+0.0052g_{pn}-0.0006g_{np}.
    \label{equ:gab}
  \end{equation}
  Here, Flambaum and Murray keep only the $g_{p}$ term. Since this
  approximation leads to a reduced sensitivity of $\kappa_{a}$ to
  $H_{\rho}^{0}$, we keep all of the terms in equation~\ref{equ:gab}.
  Assuming only contributions from isovector $\pi$ and isoscalar
  $\rho$ exchange, the dimensionless constants $g_{ab}$ can be related
  to the weak coupling constants \cite{Sus93}
  \begin{eqnarray}
    g_{pp}=g_{nn} &=& 2(\mu_{v}+2)W_{\rho}A_{\rho}H_{\rho}^{0}, \\
    g_{pn} &=& 2(2\mu_{v}+1)W_{\rho}A_{\rho}H_{\rho}^{0}
    +\sqrt{32}W_{\pi}A_{\pi}H_{\pi}^{1}, \\
    g_{np} &=& 2(2\mu_{v}+1)W_{\rho}A_{\rho}H_{\rho}^{0}
    -\sqrt{32}W_{\pi}A_{\pi}H_{\pi}^{1}, \\
    g_{p}=\frac{Z}{A}g_{pp}+\frac{N}{A}g_{pn}, \\
    g_{n}=\frac{Z}{A}g_{np}+\frac{N}{A}g_{nn}, \\
  \end{eqnarray}
  where
  \begin{eqnarray}
    A_{\rho} &=& \frac{\sqrt{2}}{Gm_{\rho}^{2}}
    \approx 0.20\times10^{6}, \\
    A_{\pi} &=& \frac{1}{Gm_{\pi}^{2}}
    \approx 4.4\times10^{6},
  \end{eqnarray}
  with $W_{\rho}\approx0.4$ and $W_{\pi}\approx0.16$.  This leads to
  the following expression for $\kappa_{a}$ in terms of $H_{\pi}^{1}$
  and $H_{\rho}^{0}$:
  \begin{equation}
    \kappa_{a}\approx1.05\times10^{5}\left(H_{\pi}^{1}
      +0.69H_{\rho}^{0}\right).
    \label{eq:kappa}
  \end{equation}
  Equation~\ref{eq:kappa} implies that it is possible to choose values
  of $H_{\pi}^{1}$ and $H_{\rho}^{0}$ which simultaneously satisfy
  both the $^{18}$F and $^{133}$Cs measurements. This solution, as can
  be seen in figure~\ref{fig:wcc}, consists of a small value for
  $H_{\pi}^{1}$ and a large value of $H_{\rho}^{0}$, at the limit of
  the DDH ``reasonable range''. In contrast, Flambaum and Murray
  assume the DDH ``best'' value of $H_{\rho}^{0}$ in their extraction
  of $H_{\pi}^{1}$.
  
\section{\boldmath Interpretation of the $^{205}$Tl Anapole Moment Measurement}
  \label{sec:tla}
  
  The $^{133}$Cs experiment is as yet the only non-zero measure of an
  anapole moment. There is, however, an earlier null measurement in
  $^{205}$Tl \cite{Vet95} which is of sufficient accuracy to be
  relevant.\footnote{There are also null results for the anapole
    moment of $^{207}$Pb \cite{Mee95,Phi96}, however the statistical
    accuracy of the measurements is not sufficient to usefully
    constrain the weak couplings.} It is interesting to compare these
  two cases.  Dmitriev and Telitsin treat both nuclei in a consistent
  way, so that we can use the two measurements as a check on the
  stability of the theory.  We analyze the $^{205}$Tl result,
  $\kappa_{a}=-0.22\pm0.30$, by the same method we used for $^{133}$Cs
  and find
  \begin{equation}
    \kappa_{a}=4.73\times10^{5}\left(H_{\pi}^{1}+0.55H_{\rho}^{0}\right).
  \end{equation}
  Because they are both odd-proton nuclei, the anapole moments of the
  two show a similar dependence on the weak coupling constants.  The
  experimental values $\kappa_{a}=-0.22\pm0.30$ for $^{205}$Tl
  \cite{Vet95} and $\kappa_{a}=0.364\pm0.062$ for $^{133}$Cs
  \cite{Fla97}, however, lead to inconsistent constraints on
  $H_{\pi}^{1}$ and $H_{\rho}^{0}$, as shown in figure~\ref{fig:wcc}.
  The inconsistency is statistically significant, 2.5$\sigma$. This
  result suggests that nuclear structure effects which are not
  included in the theory may be important in interpreting the
  measurements, assuming both measurements are correct.

\section{Discussion}
  \label{sec:dis}
  
  The relationship between the values of $H_{\pi}^{1}$ and
  $H_{\rho}^{0}$ extracted from the $^{18}$F and $^{133}$Cs
  experiments are shown in figure~\ref{fig:wcc}. Also indicated on the
  plot are the values we extract from the $^{205}$Tl experiment and
  the DDH ``best'' values and ``reasonable range''. If one assumes, as
  do Flambaum and Murray, the DDH ``best'' value of $H_{\rho}^{0}$
  value in interpreting the $^{133}$Cs result, there is indeed a
  statistically significant disagreement between the values of
  $H_{\pi}^{1}$ from the $^{133}$Cs anapole moment and from the
  $^{18}$F measurements. No model-independent determination of
  $H_{\rho}^{0}$, analogous to the method used for $H_{\pi}^{1}$ in
  $^{18}$F, is possible from nuclear processes because the one-body
  part of the corresponding potential is $\Delta I=0$. When the
  sensitivity of the anapole moment measurement to $H_{\rho}^{0}$ is
  considered, there is a solution consistent with both experiments,
  albeit an extreme one.  This solution, taken at face value, is at
  the upper limit of the DDH ``reasonable range'' and implies that
  weak hadronic interaction phenomena in nuclei are determined
  primarily by one constant, $H_{\rho}^{0}$. The apparent
  inconsistency between the $^{133}$Cs and $^{205}$Tl results,
  suggests that the nuclear theory of anapole moments is not yet
  reliable.
  
  The present controversy concerning the interpretation of measured
  nuclear anapole moments highlights the need to determine the weak
  couplings from experiments in few-nucleon systems whose
  interpretation is free from uncertainties in nuclear structure. If
  the weak meson-nucleon couplings were known, the present controversy
  could be more definitively addressed.  One would then calculate the
  values of the anapole moments using the measured values of the
  couplings from few-nucleon experiments; if the measured values of
  the anapole moments disagreed with these predictions, the problem
  would lie with the measurements, the nuclear theory, or the
  applicability of the meson-exchange model to nuclei.
  
  We now outline a technically feasible program of measurements to
  determine the most important weak meson-nucleon couplings;
  $H_{\pi}^{1}$, $H_{\rho}^{0}$, $H_{\rho}^{2}$, and the combination
  $H_{\omega}^{0}+H_{\omega}^{1}$. Experiments which measured the
  longitudinal asymmetry $A_{z}$ in \emph{pp} elastic scattering at 15
  and 45~MeV have considerably increased our knowledge of the weak
  meson-nucleon couplings \cite{Eve91,Pot74}. These experiments
  determine a linear combination of the $\Delta I=0$, 1, and 2 $\rho$
  and the $\Delta I=0$ and 1 $\omega$ couplings. For example, at
  45~MeV, $A_{z}$ is given by \cite{Ade85}
  \begin{equation}
    A_{z}=
    -0.053\left(H_{\rho}^{0}+H_{\rho}^{1}+H_{\rho}^{2}/\sqrt{6}\right)
    -0.016\left(H_{\omega}^{0}+H_{\omega}^{1}\right).
  \end{equation}
  An experiment in progress at TRIUMF \cite{Ber97} to measure $A_{z}$
  at 221~MeV is sensitive to only the $\rho$ couplings:
  \begin{equation}
    A_{z}=
    0.028\left(H_{\rho}^{0}+H_{\rho}^{1}+H_{\rho}^{2}/\sqrt{6}\right).
  \end{equation}
  Further experiments are necessary to separate the relative
  contributions of the individual couplings, $H_{\rho}^{0}$,
  $H_{\rho}^{1}$, and $H_{\rho}^{2}$. Since the $\Delta I=1$ $\rho$
  coupling is believed to be small \cite{Des80}, only one other
  independent measurement in necessary. Measurements in the \emph{np}
  system can provide the missing information.
  
  The problem of separately determining the weak $\rho$ couplings
  could be resolved by measuring the circular polarization
  $P_{\gamma}$ of the gammas emitted in the $\vec{n}+p\rightarrow
  d+\gamma$ reaction. This observable is
  primarily sensitive to the $\Delta I=0$ and 2 $\rho$ couplings
  \cite{Ade85}:
  \begin{equation}
    P_{\gamma}=0.022H_{\rho}^{0}+0.043H_{\rho}^{2}/\sqrt{6}
    -0.002H_{\omega}^{0}.
  \end{equation}
  The combination of this measurement and the TRIUMF \emph{pp}
  measurement would then independently determine $H_{\rho}^{0}$ and
  $H_{\rho}^{2}$. In practice, it is experimentally easier to measure
  the inverse reaction, the directional asymmetry in the
  photo-disintegration of the deuteron by circularly polarized
  photons, due to the low efficiency of $\gamma$-ray polarization
  analyzers. As in the case of $A_{\gamma}$, existing measurements of
  $P_{\gamma}$ (or of the inverse reaction) \cite{Lob72,Kny84,Ear88}
  are not of sufficient precision to accomplish this task. With the
  availability of very intense gamma beams from Compton
  back-scattering using free electron lasers \cite{Lit97}, a more
  precise measurement of the deuteron photo-disintegration asymmetry
  is possible and should be pursued.
    
  The problem of determining $H_{\pi}^{1}$ can be addressed by
  measuring the gamma ray asymmetry with respect to neutron spin
  direction $A_{\gamma}$ in the $\vec{n}+p\rightarrow d+\gamma$
  reaction. According to Danilov's theorem \cite{Dan65}, at thermal
  energies and below $A_{\gamma}$ is due entirely to weak processes
  that have $\Delta I=1$.  Detailed calculations \cite{Des97,Ade85}
  show that contributions from heavier mesons are suppressed and the
  process is dominated by $\pi$ exchange.  $A_{\gamma}$ is therefore
  almost completely determined by $H_{\pi}^{1}$, with other couplings
  contributing at the few-percent level \cite{Ade85}:
  \begin{equation}
    A_{\gamma}=-0.045\left(H_{\pi}^{1}-0.02H_{\rho}^{1}
      +0.02H_{\omega}^{1}+0.04H_{\rho}^{'1}\right).
  \end{equation}
  Unfortunately, existing measurements of $A_{\gamma}$
  \cite{Cav77,Alb88} are not of sufficient statistical precision
  necessary to determine $H_{\pi}^{1}$.  A precision measurement of
  $A_{\gamma}$ would be a nuclear structure independent determination
  of $H_{\pi}^{1}$, with the other couplings contributing only at the
  few-percent level. A precision measurement of $A_{\gamma}$ should be 
  pursued.
  
  Finally, we point out that additional measurements in few-nucleon
  systems, for example neutron spin rotation in H$_{2}$ or $^{4}$He
  \cite{Hec89}, would then test the applicability of the
  meson-exchange model to the weak interaction in nuclei, provided
  that the five-body calculations can be reliably performed.

\section{Summary}
  \label{sec:sum}
  
  The recent measurement of the anapole moment of $^{133}$Cs appears
  to yield a value of $H_{\pi}^{1}$ which contradicts the upper limit
  from the earlier $^{18}$F experiments. We argue that because of the
  sensitivity to $H_{\rho}^{0}$ in the $^{133}$Cs measurement, there
  is a combination of weak couplings which satisfies both experiments
  and which is (barely) in agreement with theory. In addition, the
  interpretation of the $^{133}$Cs result is inconsistent with an
  earlier null measurement in $^{205}$Tl, calling into question the
  reliability of the theory of nuclear anapole moments. All three
  experiments require assumptions about nuclear structure. We argue
  that measurements of directional asymmetry in $\vec{n}+p\rightarrow
  d+\gamma$ and in the photo-disintegration of the deuteron by
  circularly polarized photons, combined with results from \emph{pp}
  scattering, would determine the weak meson-nucleon couplings in a
  model-independent way.
  
  We would like to thank J.L. Friar, S.K. Lamoreaux, and S.I.
  Penttil{\"a} for useful discussions. This research was supported by
  the U.S.  Department of Energy under contract W-7405-ENG-36.

\begin{figure}[htbp]
  \centering
  \caption{Weak coupling constants $H_{\pi}^{1}$ and $H_{\rho}^{0}$
    extracted from the $^{18}$F (light band), $^{205}$Tl (medium
    band), and $^{133}$Cs (dark band) experiments. Also shown are the
    DDH ``best'' values (square) and ``reasonable range'' (box).}
  \label{fig:wcc}
\end{figure}

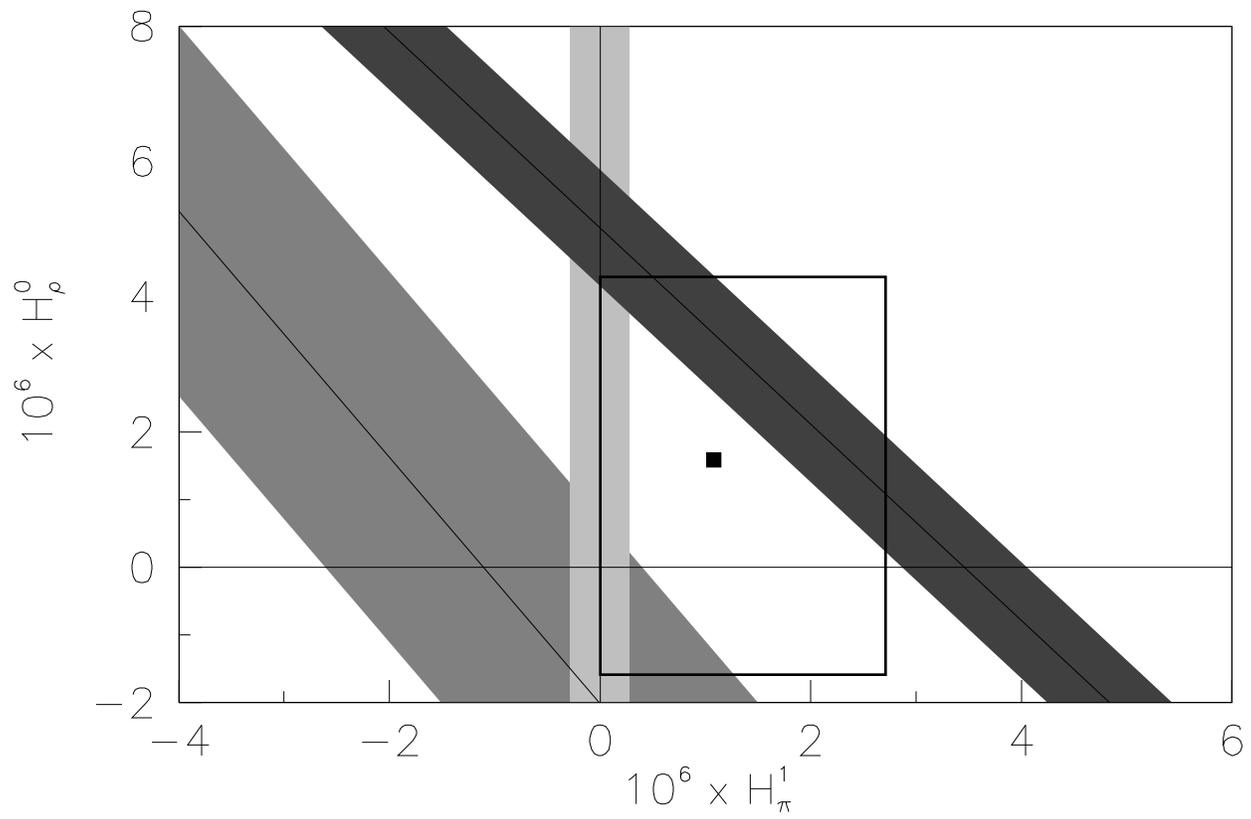
\end{document}